\documentstyle[prb,preprint,epsfig,aps]{revtex}
\begin{document}
\draft
\title{Resistance and Resistance Fluctuations in 
\\Random Resistor Networks Under Biased Percolation}
\author{C. Pennetta$^{1}$, L. Reggiani$^{1}$, Gy. Tref\'an$^2$ 
and E. Alfinito$^{1}$}
\address{$^1$ INFM - National Nanotechnology Laboratory,
Dipartimento di Ingegneria dell'Innovazione \\
Universit\`a di Lecce, Via Arnesano, I-73100 Lecce, Italy\\
$^2$ Department of Electrical Engineering, Eindhoven University 
of Technology \\5600 MB Eindhoven, The Netherlands\\
\thanks{Corresponding authors e-mail: cecilia.pennetta@unile.it }}
\date{\today}
\maketitle
\begin{abstract}
\noindent
We consider a two-dimensional random resistor network (RRN) in the presence
of two competing biased percolations consisting of the  breaking and
recovering of elementary resistors. These two processes are driven by the 
joint effects of an electrical bias and of the heat exchange with a thermal 
bath. The electrical bias is set up by applying a constant voltage or, 
alternatively, a constant current. Monte Carlo simulations are performed to 
analyze the network evolution in the full range of bias values. 
Depending on the bias strength, electrical failure or steady state are 
achieved. Here we investigate the steady-state of the RRN focusing on the
properties of the non-Ohmic regime.
In constant voltage conditions, a scaling relation is found between 
$<R>/<R>_0$ and $V/V_0$, where  $<R>$ is the average network resistance,
$<R>_0$ the linear regime resistance and $V_0$ the threshold value for the
onset of nonlinearity. A similar relation is found in constant current
conditions. The relative variance of resistance fluctuations also exhibits 
a strong nonlinearity whose properties are investigated. The power spectral 
density of resistance fluctuations presents a Lorentzian spectrum and the 
amplitude of fluctuations shows a significant non-Gaussian behavior in the 
pre-breakdown region. These results compare well with electrical breakdown 
measurements in thin films of composites and of other conducting 
materials.
\end{abstract}
%
%
\section{Introduction}
The study of electrical and mechanical stability of disordered systems 
is attracting a considerable interest in the recent literature 
\cite{hermann}$^-$\cite{pen_physd} because of its implications 
on both material technology \cite{hermann}$^-$\cite{bardhan}$^,$
\cite{scorzoni}$^-$\cite{pen_physd} and fundamental aspects related to the 
response of these systems to high external stresses 
\cite{hermann}$^-$\cite{heaney_99}$^,$\cite{stanley}$^-$\cite{sahimi}. 
Indeed, the application of a finite stress (electrical or mechanical) to a 
disordered material generally implies a nonlinear response, 
which ultimately leads to an irreversible breakdown (catastrophic behavior) 
in the high stress limit \cite{hermann}$^-$\cite{havlin}$^,$\cite{ohring}. 
Such catastrophic phenomena have been successfully studied by using 
percolation theories \cite{hermann}$^-$\cite{acharyya}$^,$
\cite{pietronero}$^-$\cite{heaney_99}$^,$\cite{pen_physd}$^-$
\cite{pen_prl_fail}. Critical phenomena near the percolation 
threshold have been widely investigated in the electrical breakdown of 
granular metals or conductor-insulator composites 
\cite{hermann}$^-$\cite{acharyya}$^,$\cite{chakrabarty}$^-$\cite{heaney_99}$^,$
\cite{stanley}$^-$\cite{pen_prl_fail}. The associated critical exponents 
have been measured \cite{hermann}$^-$\cite{havlin}$^,$
\cite{heaney,heaney_99,stauffer,sahimi,dubson89,yagil93} and theoretically 
studied using continuum or lattice percolation models 
\cite{hermann}$^-$\cite{havlin}$^,$\cite{stauffer}$^-$\cite{pen_prl_fail}.
In particular, large attention has been devoted to the critical exponents 
describing the resistance and its relative noise in terms of the medium 
properties (e.g. conducting particle fraction, defect concentration, etc.) 
\cite{hermann}$^-$\cite{heaney}$^,$\cite{heaney_99}$^,$
\cite{stauffer}$^-$\cite{pen_prl_fail}. However, very few attempts 
\cite{chakrabarty}$^-$\cite{mukherjee}$^,$\cite{pen_physc} have been made 
so far to describe the behavior of a disordered medium over the full range of 
the applied stress, i.e. by studying the response of the system to an external 
bias when the bias strength covers the full range of linear and nonlinear 
regimes. Therefore, a satisfactory understanding of breakdown phenomena 
over the full dynamical regime is still missing 
\cite{sornette96}$^,$\cite{mukherjee}. On the other hand, relevant 
information can be obtained from such a study, like: precursor phenomena, 
role of the disorder, existence of scaling laws, predictability of breakdown, 
etc. \cite{hermann}$^-$\cite{havlin}$^,$\cite{sornette96,mukherjee,stanley}.

Here we present a percolative model of sufficient generality to address the 
above issues. Our aim is to provide a theoretical framework to study response
and fluctuation phenomena under linear and nonlinear regimes in a wide class 
of disordered systems. To this purpose, we study the evolution of a random
resistor network (RRN) in which  two competing processes are present, defect 
generation and defect recovery, which determine the values of the elementary 
network resistances. Both processes are driven by the joint effect of an 
electrical bias and of the heat exchange between the network and the thermal 
bath. The bias is applied through a constant voltage or, alternatively, 
a constant current. Monte Carlo (MC) simulations are performed to investigate 
the network evolution in the full range of bias values. Depending on the bias 
strength, an irreversible failure or a stationary state of the RRN can be 
achieved. By focusing on the steady-state, we analyze the behavior of the 
average network resistance, $<R>$, and the properties of the resistance 
fluctuations as a function of the bias. In constant voltage conditions, 
a scaling relation is found between $<R>/<R>_0$ and $V/V_0$, where $<R>_0$ 
is the linear regime resistance and $V_0$ the threshold value for the onset 
of nonlinearity. A similar relation is found in constant current conditions. 
The relative variance of resistance fluctuations also exhibits a strong 
nonlinearity in the pre-breakdown regime whose properties are discussed. 
The fluctuation analysis is completed by a further investigation of the noise 
resistance spectrum and of the Gaussian features of the fluctuation amplitudes.
Theoretical results agree with electrical breakdown measurements 
in thin films of composites \cite{nandi,mukherjee} and of other conducting 
\cite{scorzoni,bloom,seidler} or insulating materials \cite{vandewalle}. 

The paper is organized as follows. In Sect. 2 we briefly describe the model 
used. Section 3 presents the results of the MC simulations for the resistance
and its fluctuations. The main conclusions are drawn in Sect. 4.
\section{Model}
We study a two-dimensional random resistor network of total resistance $R$, 
made of $N_{tot}$ resistors, each  of resistance $r_n$, disposed on a square 
lattice. We take a square geometry, $N \times N$, where $N$ determines the 
linear size of the lattice, with the total number of resistors being 
$N_{tot} = 2 N^2$. For the comparison with resistivity measurements in thin 
films, the value of $N$ can be related to the ratio between the size of the 
sample and the average size of the grains composing the sample. 
An external bias, represented by a constant voltage $V$ or by a constant 
current $I$, is applied to the RRN through electrical contacts realized by 
perfectly conducting bars at the left and right hand sides of the network. 
A current $i_n$ is then flowing through each resistor. The RRN interacts with
a thermal bath at temperature $T_0$ and the resistances $r_n$ are taken to 
depend linearly on the local temperatures, $T_n$, as:
\begin{equation}
r_{n}(T_{n})=r_{0}[ 1 + \alpha (T_{n} - T_0)]
\label{eq:tcr}
\end{equation}
In this expression $\alpha$ is the temperature coefficient of the resistance 
and $T_n$ is calculated by adopting a biased percolation model 
\cite{pen_mcs,pen_prl_fail} as:
\begin{equation}
T_{n}=T_{0} + A \Bigl[ r_{n} i_{n}^{2} + {B \over N_{neig}}
\sum_{l=1}^{N_{neig}}  \Bigl( r_{l} i_{l}^2   - r_n i_n^2 \Bigr) \Bigr]
\label{eq:temp}
\end{equation}
Here, $N_{neig}$ is the number of first neighbours around the n{\em th}
resistor, the parameter $A$, measured in (K/W), describes the heat coupling
of each resistor with the thermal bath and it determines the importance of
Joule heating effects. The parameter $B$ is taken to be equal to $3/4$ to 
provide a uniform heating in the perfect network configuration.
We note that Eq.~(\ref{eq:temp}) implies an instantaneous thermalization of
each resistor at the value $T_n$, therefore, by adopting  Eq.~(\ref{eq:temp}),
for simplicity we are neglecting  time dependent effects which are discussed
in Ref. [\onlinecite{sornette92}].

In the initial state of the network (corresponding to the perfect network
configuration with no heating) all the resistors are identical: 
$r_n \equiv r_0$. Now, we assume that two competing percolative processes 
act to determine the RRN evolution. The first process consists of generating
fully insulating defects (resistors with very high resistance, i.e. broken
resistors) with probability \cite{pen_mcs,pen_prl_fail} 
$W_D=\exp( -E_D/K_B T_n )$, where $E_D$ is an activation energy 
characteristic of the defect and $K_B$ the Boltzmann constant.
The second  process consists of recovering the insulating defects
with probability  $W_R=\exp (-E_R/K_B T_n )$, where $E_R$ is an activation 
energy characteristic of this second process. For $A \neq 0$, 
Eq.~(\ref{eq:temp}) implies that both the processes (defect generation and 
defect recovery) are correlated percolations. Indeed, the probability of
breaking (recovering) a resistor is higher in the so called "hot spots" of
the RRN \cite{stauffer}. On the other hand, for $A=0$ Eq.~(\ref{eq:temp}) 
yields $T_n \equiv T_0$, which corresponds to random percolations 
\cite{stauffer,pen_prl_stat}. The same is true for vanishing small bias 
values, when Joule heating effects are negligible. 

As a result of the competition between these two percolations and depending 
on the parameters related to the particular physical system ($E_D$, $E_R$, 
$A$, $\alpha$, $r_0$, $N$) and on the external conditions (specified by the 
bias conditions and the bath temperature), the RRN reaches a steady 
state or exhibits an irreversible breakdown. In the first case, the network 
resistance fluctuates around an average value $<R>$. In the second case, 
a critical fraction of defects $p_c$, corresponding to the percolation 
threshold, is reached, i.e. $R$ diverges due to the existence of at least 
one continuous path of defects between the upper and lower sides of the 
network \cite{stauffer}. We note that in the limit of a vanishing bias 
(random percolation) and infinite lattices ($N \rightarrow \infty$), 
the expression: $E_R < E_D + K_BT_0 \, \ln [1 + \exp(-E_D/K_BT_0)]$ provides a 
sufficient condition for the existence of a steady state \cite{pen_prl_stat}.

The evolution of the RRN is obtained by MC simulations carried out according 
to the following iterative procedure. (i) Starting from the perfect lattice
with given local currents, the local temperatures $T_n$ are calculated 
according to Eq.~(\ref{eq:temp}); (ii) the defects are generated with 
probability $W_D$ and  the resistances of the unbroken resistors are changed 
as specified by Eq.~(\ref{eq:tcr}); (iii) the currents $i_n$ are calculated 
by solving Kirchhoff's loop equations by the Gauss elimination method and the
local temperatures are updated; (iv) the defects are recovered with 
probability $W_R$ and the temperature dependence of unbroken resistors is 
again accounted for; (v) $R$, $i_n$ and $T_n$ are finally calculated and the 
procedure is iterated from (ii) until one of the two following possibilities 
is achieved. In the first, the percolation threshold is reached. In the 
second, the RRN attains a steady state; in this case the iteration runs 
long enough to allow a fluctuation analysis to be carried out.   
Each iteration step can be associated with an elementary time step on an 
appropriate time scale (to be calibrated with experiments). In this manner 
it is possible to represent the simulation of the resistance evolution over 
either a time or a frequency domain according to convenience.

In the simulations, as reasonable values of the parameters, we have taken:
$N =75$, $r_0=1 \ (\Omega)$, $\alpha = 10^{-3}$ (K$^{-1}$),
$A=5 \times 10^5$ (K/W), $E_D = 0.17$ (eV),  $E_R = 0.043$ (eV),
and $T_0=300$ (K) if not stated otherwise. The values of the external bias 
range from $0.05 \le I \le 2.8$ (A) under constant current conditions, and 
from $0.05 \le V \le 3.5$ (V) under constant voltage conditions.
\section{Results}
Figures 1 (a) and 1 (b) report a sampling of the resistance evolutions coming
out from the simulations under steady state conditions for the cases of
constant voltage (Fig. 1 (a)) and constant current (Fig. 1 (b)), respectively.
At increasing the external bias, a remarkable increase of the average 
resistance and of the amplitude of fluctuations is evident in both cases
while approaching the breakdown. This is reached by applying an external bias
just above the highest values shown in these figures. Precisely, 
the RRN becomes unstable under constant voltage already by applying a 
voltage just above $3.0$ (V) and, under constant current, by applying a 
current just above $2.1$ (A).

In the following, Figs. 2 to 4 will detail the behavior of the average 
resistance while Figs. 5 to 10 will focus on the results of resistance 
fluctuations.
\subsection{Resistance}
Figure 2 reports the average value of the RRN resistance as a function
of the applied bias (current or voltage). Each value is calculated by 
considering the time average on a single steady-state realization and then 
averaging over 20 independent realizations (RRNs subjected to the same bias 
conditions). The numerical uncertainty is found to be within 0.01\% at worst.
At the lowest biases the resistance takes a value $<R>_0$ which represents 
the intrinsic linear response property of the network (Ohmic regime) 
\cite{pen_prl_stat}. We note that for the activation energy values here 
chosen, the average fraction of defects at the lowest bias $<p>_0$ is very 
small ($<p>_{0}=0.545 \times 10^{-2}$) so that $<R_0>=0.9978$ ($\Omega$) 
is about 1\% greater than the perfect network resistance.
Under constant current conditions, when the current overcomes a certain value,
$I_0$, the average resistance starts to become dependent on the bias. 
Thus, $I_0$ sets the current scale value for the onset of nonlinearity.
By adopting as criterium for the onset of nonlinearity a resistance 
increase of $0.05 \%$ over the linear response value, we determine
$I_0=0.090 \ \pm \ 0.005$ (A). For currents above $I_0$ the resistance 
increase is smooth at moderate bias and it exhibits a sharp ramp (typical 
of a catastrophic behavior) at high bias, until a threshold current value
$I_b=2.10$ (A) above which the RRN undergoes an irreversible breakdown.
A step of current values $\delta I=0.05$ (A) has been used to determine 
$I_b$, thus this value of $\delta I$ gives the maximum uncertainty on $I_b$.
Accordingly, we have found $I_b/I_0=23 \pm 1$ and $<R>_b/<R>_0=1.6 \pm 0.1$, 
where $<R>_b$ is the last stable value of the resistance calculated before 
the breakdown. We remark that the uncertainty on the ratio $<R>_b/<R>_0$ is 
mainly due to the uncertainty on $<R>_b$ which reflects, amplified by the 
nonlinearity, the uncertainty on $I_b$. Similarly, under constant voltage 
conditions we have found $V_b/V_0=32 \pm 2$ and $<R>_b/<R>_0=1.5 \pm 0.1$, 
where $V_0=0.095  \ \pm \ 0.005$ (V) and $V_b=3.00$ (V) are the voltage values 
corresponding respectively to the nonlinearity onset and to the electrical 
breakdown. The maximum uncertainty on $V_b$ is $\delta V = 0.05$ (V). 
The significantly higher value of the ratio  $V_b/V_0$, when compared with 
that of the ratio $I_b/I_0$, is a quantitative indication that the system 
is more robust when biased under constant voltage than under constant current 
conditions. This property is further emphasized by the fact that the increase 
of the resistance in the pre-breakdown region exhibits a lower slope under 
constant voltage than under constant current conditions. 
It must be noticed that in spite of the significant difference of the ratios
$V_b/V_0$ and $I_b/I_0$, the ratio $<R>_b/<R>_0$ remains the same, within the 
error, under the different bias conditions. This result agrees with 
measurements of the ratio $<R>_b/<R>_0$ performed in composites 
under the Joule regime \cite{mukherjee}.  

To better analyze the dependence on the bias of the average resistance, 
Figs. 3 (a) and 3 (b) report the log-log plot of the relative variation 
of the resistance as a function of $V/V_0$ and $I/I_0$, respectively. 
Figure 3 (a) shows that the relative 
variation of the average resistance scales with the ratio $V/V_0$ in the 
whole region of applied voltages up to breakdown as: 
\begin{equation}
{<R>_V \over <R>_0} = 1 + a \left ({V \over V_0} \right )^{\theta}
\label{eq:res_v}
\end{equation}
where $a=(1.1 \pm 0.1) \times 10^{-4}$ is a dimensionless coefficient and 
$\theta=2.1 \pm 0.1$. Figure 3 (b) shows that also under constant current 
the relative variation of the average resistance scales with the ratio 
$I/I_0$ in the moderate bias region, as:
\begin{equation}
{<R>_I \over <R>_0} = 1 + a' \left ({I \over I_0} \right )^{\theta}
\label{eq:res_i}
\end{equation}
with $a'=(3.8 \pm 0.1) \times 10^{-4}$ and the same exponent $\theta$ of 
Eq.~(\ref{eq:res_v}), within numerical uncertainty. However, in this case 
we notice that in the pre-breakdown region the relative variation of the 
average resistance exhibits a superquadratic behavior characterized by a 
power law $(I/I_0)^{\theta_I}$ with ${\theta_I}=3.7 \pm 0.1$.

This behavior of the average resistance with the applied bias can be 
understood as follows. From a macroscopic point of view, in a degradation 
process associated with a resistance increase, constant current conditions 
lead, at increasing bias, to a superquadratic increase of the power dissipated
through Joule heating and thus to a major efficiency in the defect generation.
By contrast, constant voltage conditions lead, at increasing bias, to a growth
 of the dissipated power which is subquadratic in the applied voltage,
thus implying a minor efficiency in the generation of defects. 
In this respect, it must be underlined that the resistance increase driven by 
both kind of bias is further enhanced by the positive value of the temperature
coefficient of the resistance.

From a microscopic point of view, the RRN resistance depends on the average 
fraction of defects, $<p>$, according to the expression \cite{stauffer}:
\begin{equation}
<R> \sim |<p> -\, p_c \ |^{-\mu} 
\label{eq:res_scal}
\end{equation}
where, for biased percolation, $p_c$ and $\mu$ are functions of the bias 
strength \cite{pen_mcs,pen_prl_fail}. Of course, for vanishing bias, $p_c$ 
and $\mu$ take the well known values corresponding to random percolation: 
$p_c=0.5$ (for a square lattice) and $\mu=1.303$ (universal value) 
\cite{stauffer}. On the other hand, the dependence of $p_c$ and $\mu$ 
on the bias makes the analysis of $<R>$ in terms of Eq.~(\ref{eq:res_scal}) 
quite problematic. Nevertheless, it is interesting to analyze the behavior of 
$<p>$ on the bias strength and, to this purpose, Fig. 4 reports on a log-log 
plot the relative variation of $<p>$ as a function of the normalized bias. 
As evidenced by the figure, for the case of constant voltage conditions the 
average fraction of defects scales with the same exponent $\theta$ over the 
full range of bias values, as:
\begin{equation}
{<p>_V \over <p>_0} = 1 + b \left ({V \over V_0} \right )^{\theta}
\label{eq:pscal_v}
\end{equation}
with $b=(5.8 \pm 0.3) \times 10^{-3}$ a dimensionless coefficient. 
A similar expression holds under constant current conditions at moderate bias:
\begin{equation}
{<p>_I \over <p>_0} = 1 + b' \left ({I \over I_0} \right )^{\theta}
\label{eq:pscal_i}
\end{equation}
with $b'=(5.3 \pm 0.3) \times 10^{-3}$. However, under constant current, 
there is a significant increase of the slope in the pre-breakdown region 
up to a nearly above cubic power law. 

The behavior of the average fraction of defects in the moderate bias region 
can be understood by generalising the results obtained for random 
stationary percolations \cite{pen_prl_stat} to the presence of a moderate 
external bias responsible of Joule heating. In fact, for a RNN in a stationary
state resulting from the competition of two random percolations, it has been 
found \cite{pen_prl_stat} that $<p>=x/(1+x)$, with 
$x=W_{0D}(1-W_{0R})/W_{0R}$, where $W_{0D}$ and $W_{0R}$ are the probabilities
defined in section II with $T_n \equiv T_0$. At moderate bias, when the defect distribution is rather 
homogeneous and the variation of $<R>$ small, we can take $x=x(T)$ with 
$T=T_0+\Delta T$, where in the spirit of a mean-field theory, the average 
temperature increase can be expressed as $\Delta T = A<R>_0 I^2/(2N^2)$. 
Therefore, by differentiating:
\begin{equation}
\Delta <p> = {1 \over (1+x)^2_0} \left ({dx \over dT} \right )_0 \Delta T
\label{eq:delta_p}
\end{equation}
which by simple manipulation gives:
\begin{equation}
{\Delta <p> \over <p>_0} =
\left [ {A<R>_0 I_0^2 \over 2 N^2 <p>_0 [(1+x)^2]_0} 
\left ( {dx \over dT} \right)_0 \right ] \left ({I \over I_0} \right )^2 
\label{eq:rel_p}
\end{equation}
The term in the squared brackets of Eq.~(\ref{eq:rel_p}) can thus be 
identified with the $b'$ coefficient in  Eq.~(\ref{eq:pscal_i}). In fact,
by introducing the parameters used in the simulations we find for the squared 
brackets in Eq.~(\ref{eq:rel_p}) the value $5.5 \times 10^{-3}$ which well 
agrees with the previous reported value of $b'$. For constant voltage the 
factor $A<R>_0 I_0^2$ inside Eq.~(\ref{eq:rel_p}) must be replaced with  
$AV_0^2/<R>_0$. This gives for the square bracket in Eq.~(\ref{eq:rel_p})
the value of $6.1 \times 10^{-3}$, which again well agrees with the value 
of $b$ in Eq.~(\ref{eq:pscal_v}). Therefore, at moderate bias, 
the quadratic behavior of $<p>$ is well explained by Eq.~(\ref{eq:rel_p}). 
In the pre-breakdown region, under constant current conditions, the 
significant increase of the resistance and the defect filamentation 
characteristic of biased percolation \cite{pen_mcs,pen_prl_fail} add to 
give a superquadratic increase of $<p>$ with the bias. By contrast, under 
constant voltage, these two effects act in an opposed manner, thus their 
compensation results in a quadratic dependence of $<p>$ over the full 
range of bias values.

From the above behavior of the average defect fraction we can now understand
the bias dependence of $<R>$. At moderate bias ($I/I_0$ or $V/V_0 \le 10$) 
the increase of $<R>$ is rather small. Accordingly, super and subquadratic 
effects on the dissipated power are negligible and the average fraction of 
defects over the intrinsic value grows quadratically with the bias strength 
(Eq.~(\ref{eq:rel_p})); at these bias values  $<p> \ \ll \ 1$  thus 
$<R> \ \propto \ <p>$ and also the average resistance
grows quadratically with bias. At high bias (i.e. $I/I_0$ or $ V/V_0 > 10$), 
when entering the pre-breakdown region, the increase of $<R>$ is significant. 
Accordingly, super and subquadratic effects on the dissipated power are 
relevant as well as the increase of the average fraction of defects which 
now takes a nonuniform distribution (biased percolation). As a consequence, 
the percolation threshold is lowered with respect to the random percolation 
value \cite{pen_mcs}. Depending on constant current or constant voltage 
conditions, this biased percolation effect adds to further enhancing the 
increase of the resistance well above the quadratic law or to keep the 
quadratic behavior, respectively.

This interpretation is confirmed by MC simulations performed with the same 
parameters but setting the temperature coefficient of the resistance 
$\alpha=0$. In this case, we have found that the onset of nonlinearity as 
well as the electrical breakdown occur at higher values of the bias. 
Precisely we have found $I'_0=0.25 \pm 0.03$ (A), $I'_b=2.65 \pm 0.05$ (A) 
(thus $I'_b/I'_0=11 \pm 1$), $<R>_b/<R>_0 = 1.2 \pm 0.1$ under constant 
current and $V'_0=0.25 \pm 0.03$ (V), $V'_b=3.25 \pm 0.05$ (V) 
(thus $V_b/V_0=13 \pm 1$) and $<R>_b/<R>_0 = 1.3 \pm 0.1$ under constant 
voltage conditions. The quadratic regime is always present at moderate bias 
values. However, in the pre-breakdown regime we have found 
${\theta'_I}=3.1 \pm 0.1$ for constant current and ${\theta'_V}=2.8 \pm 0.1$ 
for constant voltage conditions, respectively. 

We finally conclude this section by remarking that the behavior described by 
Eqs.~(\ref{eq:res_v}) and ~(\ref{eq:res_i}) well agrees with measurements of 
the resistance of composites made in the Joule regime up to breakdown 
\cite{mukherjee}.
\subsection{Fluctuations}
It is well known that important information about the properties
and the stability of different  systems (physical, biological, social, etc.) 
can be obtained by studying the fluctuations around the average value of 
some characteristic quantities of the system 
\cite{hermann}$^-$\cite{havlin}$^,$\cite{nandi}$^-$\cite{bloom}$^,$
\cite{rammal85}$^-$\cite{yagil93}$^,$\cite{kolek}$^-$\cite{weissman}. 
Several features of the fluctuations are usually considered, and among these, 
the most important are the variance (or the relative variance), the power 
spectrum and the Gaussianity property. Here, we analyze the fluctuations 
of the RNN resistance in the full range of bias values, devoting particular 
attention to the features that can be identified as failure precursors.
Figure 5 reports the relative variance of resistance fluctuations, 
$\Sigma \equiv <\delta R^2> / <R>^2$, as a function of the applied bias.
The data correspond to the same simulations presented in Fig. 2.
The same procedure of time averaging over a single simulation and then
making ensemble averages over $20$ realizations provides an
uncertainty of $ 3\%$ at worst. At low bias, the relative variance is found 
to achieve a constant value $\Sigma_0 = (4.0 \pm 0.1) \times 10^{-6}$.
This value represents an intrinsic property of the system determined
by the competition of two random percolations, as already discussed
for the case of the resistance. A detailed analysis of the scaling 
properties of $\Sigma_0$ can be found in Ref. [\onlinecite{pen_prl_stat}].
At increasing bias, when $I>I_0$ or $V>V_0$,  the systematic increase
of $\Sigma$ reveals the existence of a nonlinear regime of the response.
Similarly to the behavior of the resistance, by approaching the electrical
breakdown $\Sigma$ exhibits a significant increase which is steeper
under constant current than under constant voltage conditions.
This feature is ascribed to the better stability of the RRN under constant 
voltage, since resistance fluctuations in excess over the mean value are 
damped in this condition. By contrast, the same kind of fluctuations are 
enhanced under constant current conditions. To detail the dependence of 
$\Sigma$ on the bias, we have reported in Fig. 6 (a) its relative variation,
$[\Sigma_V-\Sigma_0]/\Sigma_0$, as a function of $V/V_0$ under constant 
voltage conditions. Two regions can be identified in the nonlinear regime: 
a moderate bias region ($V/V_0 < 10$), characterized by a quadratic dependence
on the applied voltage, and a pre-breakdown region at the highest voltages 
($V/V_0 > 10$), where a superquadratic dependence up to a cubic power law is 
evidenced. Accordingly, in the first region it is:
\begin{equation}
{\Sigma_V \over \Sigma_{0}} = 1 + c \left ({V \over V_0} \right )^{2}
\label{eq:sigm_v}
\end{equation}
with $c = (1.14 \pm 0.06) \times 10^{-2}$ a dimensionless coefficient, while 
in the second region we have found:
\begin{equation}
{\Sigma_V \over \Sigma_{0}} \sim \left ({V \over V_0} \right )^{\eta_V}
\label{eq:sigm_vb}
\end{equation}
where $\eta_V = 3.0 \pm 0.1$. This behavior of $\Sigma$ under constant voltage
conditions should be compared with that of the average resistance which 
remains always a quadratic function up to the breakdown, as shown 
in Fig. 3 (a). The emergence of a superquadratic behavior of $\Sigma$ in the 
pre-breakdown region reflects the higher sensitivity of the resistance 
fluctuations with respect to the average value of the resistance to the 
instability of the network. The same analysis performed for the case of 
constant current is shown in Fig. 6 (b). Again, at moderate bias 
$[\Sigma_I-\Sigma_0]/\Sigma_0$ increases quadratically with the current as:  
\begin{equation}
{\Sigma_I \over \Sigma_{0}} = 1 + c' \left ({I \over I_0} \right )^{2}
\label{eq:sigm_i}
\end{equation}
with $c' = (6.6 \pm 0.4) \times 10^{-3}$ a dimensionless coefficient. 
On the other hand, the superquadratic dependence characterizes the 
pre-breakdown region as:
\begin{equation}
{\Sigma_I \over \Sigma_{0}} \sim \left ({I \over I_0} \right )^{\eta_I}
\label{eq:sigm_ib}
\end{equation}
with $\eta_I=5.4 \pm 0.1$. We note, that $\eta_I$ is significantly greater 
than $\eta_V$, according to the behaviors shown in Fig. 5. Moreover,  
$\eta_I$ is greater than the resistance exponent in the pre-breakdown region 
$\theta_I$. It is noteworthy, that by neglecting the effect of the temperature
coefficient of the resistance, i.e. by taking $\alpha=0$, the moderate bias 
region remains characterized by a a quadratic increase which is common 
to both constant current and constant voltage conditions. However, in the 
pre-breakdown region we have found $\eta_V=\eta_I=4.0 \pm 0.1$. Thus,
the above results prove that the quadratic dependence of $\Sigma$ on the bias
in the moderate bias region is a feature independent of the conditions on 
which the bias is applied and also independent of the value of the thermal 
coefficient of the resistance. This reflects the fact that at moderate bias 
the two following conditions are satisfied: i) the variation of the average 
fraction of defects follows Eq.~(\ref{eq:rel_p}); ii) the average fraction of 
defects is much smaller than the percolation threshold value, 
i.e. $<p> \ \ll \ p_c$. 
Therefore, in this nearly perfect regime 
$\Sigma \ \propto \ <p>\ \propto \ <R>$ \cite{pen_prl_stat}. On the other 
hand, the superquadratic behavior of $\Sigma$ in the pre-breakdown region is a 
consequence of the fact that at high biases the local correlations in the 
defect generation and recovery processes become strong. Defect filamentation, 
characteristic of biased percolation \cite{pen_mcs,pen_prl_fail}, implies that
a small variation in the number of defects bring strong resistance 
fluctuations.

Figure 7 reports the relative variance of resistance fluctuations normalized 
to the linear regime value as a function of the normalized resistance. 
We can see that the former quantity exhibits an increase over nearly two 
orders of magnitude in the same range of bias where the latter quantity 
increases for about $60 \%$ only. The log-log plot of 
$[\Sigma - \Sigma_0]/\Sigma$ as a function of $[<R>-<R>_0]/<R>_0$, shown in 
Fig. 8, confirms the proportionality of $\Sigma$ with  $<R>$ 
in the moderate bias region and evidences the following power-law in the 
pre-breakdown region. 
\begin{equation}
{[\Sigma - \Sigma_0] \over \Sigma_0} \sim {[<R> - <R>_0] \over <R>_0}^{\zeta}
\end{equation}
where the value of the exponent, $\zeta = 1.5 \pm 0.1$, within the statistical
uncertainty is the same for constant voltage or constant current conditions. 
This result is consistent with Eqs.~(\ref{eq:res_v}) and (\ref{eq:res_i}).

To complete the study of resistance fluctuations, we have investigated the
Gaussian properties of the fluctuation amplitudes and the spectra in the 
frequency domain. Figures 9 (a)  and 9 (b) report the distribution function
of the resistance fluctuations $p(R)$ for different currents and different
bath temperatures, respectively. The dotted and the long-dashed curves 
in Fig. 9 (a) correspond to fit with a Gaussian distribution the data 
corresponding to different current values in the pre-breakdown region
at  $T_0=300$ (K), while in Fig. 9 (b) to fit the data corresponding to 
different bath temperatures at $I=1.5$ (A). When approaching the breakdown 
conditions, at increasing current in Fig. 9 (a) and at increasing temperature 
in Fig. 9 (b), the simulations show the onset of a non-Gaussian behavior 
characterized by the enhancement of the probability for the positive 
fluctuations with respect to the Gaussian distribution. The emergence of 
a non-Gaussian behavior near the breakdown, in agreement with experiments 
\cite{weissman}$^,$\cite{seidler,vandewalle}, can be considered a relevant 
precursor of  failure. 

Figures 10 (a) and 10 (b) show the spectral densities of resistance 
fluctuations for the same conditions of  Fig. 9 (a) and 9 (b). The spectral 
densities have been calculated by Fourier transforming the corresponding
correlation functions $C_{\delta R}(t)$. We have found Lorentzian spectra in 
all the cases. For a given temperature,  within the numerical uncertainty, 
the corner frequency is found to be independent of the applied current, while 
the value of the plateau increases more than quadratically with the current 
in the pre-breakdown region. For a given current in the pre-breakdown region, 
both the corner frequency and the value of the plateau are found to increase 
at increasing temperatures. This fact indicates that the characteristic
times of fluctuations while depending on temperature they are independent 
of the applied bias. We finally remark that the power spectral density in 
Figs. 10 (a) and 10 (b) are in good agreement with experiments 
\cite{scorzoni,ohring}$^,$\cite{weissman}.

\section{Conclusions}
We have studied the stationary state of a two-dimensional RRN resulting
from the competition between two biased percolations. The two percolative 
processes consist of the breaking and recovering of elementary resistors and  
they are driven by the joint effects of an electrical bias and of the heat 
exchange with a thermal bath. The electrical bias is set up by applying a 
constant voltage or a constant current.
MC simulations have been performed to analyze the network resistance  
and its fluctuation properties in the full range of bias values, 
covering linear and nonlinear regimes up to the breakdown limit. 
The nonlinear regime starts for biases greater than the threshold values, 
$V_0$ or $I_0$ (nonlinearity onset values), and it extends until the values 
$V_b$ or $I_b$ (breakdown values). The ratios $V_b/V_0$, $I_b/I_0$, 
$<R>_b/<R>_0$ can be considered as relevant indicators characterizing 
the breakdown properties of the system. We have found that the ratio 
$<R>_b/<R>_0$ is independent of the bias conditions but it depends on the
thermal coefficient of the resistance. This result agrees with measurements
of this ratio performed in composites \cite{mukherjee}. 
Moreover, we have found that under constant voltage conditions the relative 
variation of the average resistance scales quadratically with the ratio 
$V/V_0$ over the full nonlinear regime. A similar relation has been found 
under constant current conditions, but in this case a superquadratic 
dependence emerges in the pre-breakdown region ($ I> 10 \, I_0$). 
For what concerns the relative variance of resistance fluctuations 
we have found that at moderate bias ($I_0 < I < 10 \, I_0$) it 
grows quadratically with the bias, independently of constant-current or 
constant-voltage conditions. On the other hand a superquadratic dependence 
appears in the pre-breakdown region. The presence of two distinct regions in
the nonlinear regime has been explained in the following terms. 
At moderate bias there is a small Joule heating giving rise to the 
competitions of two nearly random percolations and to a small variation of 
the RRN resistance. At high bias, local correlations typical of biased 
percolation become important. Moreover, also the resistance variation 
becomes strong. Therefore, as a consequence of both these effects the 
resistance noise depends superquadratically on the bias. The power spectral 
density of resistance fluctuations is found to exhibit a Lorentzian spectrum 
and in the pre-breakdown region the amplitude of fluctuations shows the onset 
of non-Gaussian behaviors. Theoretical results agree qualitatively, and in 
some cases quantitatively, with breakdown experiments performed in 
composites \cite{nandi}$^,$\cite{mukherjee} and in other conducting 
\cite{scorzoni,bloom}$^,$\cite{seidler} or nonconducting materials 
\cite{vandewalle}. We conclude that the study of the stationary state of a 
RRN, resulting from the competition of biased percolations, provides an 
unifying framework for the interpretation of several non-linear transport 
phenomena in a variety of disordered systems. 

\vspace{0.15 cm}\noindent
ACKNOWLEDGEMENTS
\vskip 0.5 truecm\noindent
This research is performed within the STATE project of INFM.
Partial support is also provided by the MADESS II project of the Italian 
National Research Council and the ASI project, contract I/R/056/01. 
We thank Prof. L.B. Kish and Dr. Z. Gingl who introduced us to the percolation
theory.

\newpage\noindent

\newpage\noindent
FIGURE CAPTIONS
 
\vskip1pc\noindent
Fig. 1 (a) -  Resistance evolutions of a steady state $75\times 75$
network under constant voltage conditions for increasing values of the bias. 
Going from the lower to the upper curves, the voltage is $V = 0.2$, $0.5$, 
$1.0$, $1.5$, $2.0$, $3.0$ (V). The values of all the other parameters are 
specified in the text. 

\vskip1pc\noindent
Fig. 1 (b) - Resistance evolutions of a steady state $75\times 75$ network 
under constant current conditions for increasing values of the bias. 
Going from the lower to the upper curves, the current is:  
$I = 0.2$, $0.5$, $1.0$, $1.5$, $2.1$ (A). The values of all the other 
parameters are specified in the text.

\vskip1pc\noindent
Fig. 2 - Average network resistance as a function of the external bias.
The averages are calculated over $20$ networks of sizes $75 \times 75$ 
subjected  to the same bias conditions. The circles correspond to biasing 
the network by an external constant current, the triangles correspond to a
constant voltage bias.

\vskip1pc\noindent
Fig. 3 (a) - Log-Log plot of the relative variation of the network resistance 
calculated under constant voltage as a function of the ratio $V/V_0$. 
The data are the same of those in Fig. 2 and only the nonlinear regime 
is shown. The dotted line represents the power-law fit with a value for 
the exponent $\theta = 2.1 \pm 0.1$.

\vskip1pc\noindent
Fig. 3 (b) -  Log-Log plot of the relative variation of the network resistance 
calculated under constant current as a function of the ratio $I/I_0$.
The data are the same of those in Fig. 2 and only the nonlinear regime is 
shown. The long-dashed line represents the power-law fit of the data in the 
moderate bias region, the value of the exponent is $\theta = 2.1 \pm 0.1$. 
The solid line fits the data in the pre-breakdown region with a power-law, 
the value of the exponent is $\theta_I = 3.7 \pm 0.1$.

\vskip1pc\noindent
Fig. 4 -  Log-Log plot of the relative variation of the average number of 
defects under constant voltage (triangles) and under constant current 
conditions (open circles) as a function of the normalized bias ($V/V_0$ or
$I/I_0$). The solid line represents the power-law fit with a value for the 
exponent $\theta = 2.1 \pm 0.1$. The long-dashed line fits
constant-current data in the pre-breakdown region with a power-law of 
exponent $3.3 \pm 0.1$.

\vskip1pc\noindent
Fig. 5 - Relative variance of resistance fluctuations, $\Sigma$, under 
constant voltage conditions (triangles) and under constant current conditions
(open circles) as a function of the external bias. Each point has been 
obtained by averaging the variances calculated for $20$ networks of sizes 
$75 \times 75$ subjected to the same bias conditions. 

\vskip1pc\noindent
Fig. 6 (a) - Log-Log plot of the relative variation of $\Sigma$ under constant 
voltage conditions as a function of $V/V_0$. The long-dashed line represents 
the power-law fit to the data in the moderate bias region, the value of the 
exponent is $2.0 \pm 0.1$. The solid line fits data in the pre-breakdown 
region with a power-law of exponent $3.0 \pm 0.1$.

\vskip1pc\noindent
Fig. 6 (b) - Log-Log plot of the relative variation of $\Sigma$ under constant 
current conditions as a function of $I/I_0$. The long-dashed line represents 
the power-law fit to the data in the moderate bias region, the value of the 
exponent is $2.1 \pm 0.1$. The solid line fits data in the pre-breakdown 
region with a power-law of exponent $5.4 \pm 0.1$.

\vskip1pc\noindent
Fig. 7 - Relative variance of resistance fluctuations normalized to the 
linear regime value as a function of the normalized average resistance. 
Open circles refer to constant current  condition and triangles to constant
voltage condition, respectively. 

\vskip1pc\noindent
Fig. 8 - Log-Log plot of the relative variation of $\Sigma$ as a function of 
the relative variation of the average resistance under constant voltage 
condition (open circles) and constant voltage condition (triangles).
The long-dashed line gives the best fit with a power-law with 
exponent $1.0 \pm 0.1$. The solid and the dotted lines fit with a power-law 
the data in the pre-breakdown region, the exponent are $1.5 \pm 0.1$ and 
$1.6 \pm 0.1$, respectively.  

\vskip1pc\noindent
Fig. 9 (a) - Distribution function of the resistance fluctuations for two
values of the applied current. Open circles refer to $I=1.5$ (A) and full 
diamonds to $I=1.8$ (A). The scale is a linear-log, therefore the dashed and 
the long-dashed curves, coming from parabolic fits of the two sets of data,
correspond to Gaussian distributions. 

\vskip1pc\noindent
Fig. 9 (b) - Distribution function of the resistance fluctuations under 
constant current conditions for two values of the substrate temperature. 
Open circles refer to $T_0=300$ (K) (the same data reported in Fig. 9 (a) ) 
and full triangles to $T_0=450$ (K). The scale is linear-log, therefore 
the dashed and the long-dashed curves, coming from parabolic fits of the 
two sets of data, correspond to Gaussian distributions.

\vskip1pc\noindent
Fig. 10 (a) - Power spectral density of resistance fluctuations under constant
current conditions for $I=1.5$ (A) (dotted curve) and for $I=1.8$ (A) (solid
curve). In both cases the substrate temperature is $300$ (K).

\vskip1pc\noindent
Fig. 10 (b) - Power spectral density of resistance fluctuations under constant
current condition for $T_0=300 \ K$ (dotted curve) and for $T_0=450$ (K) 
(solid curve). In both cases the bias current is $1.5$ (A).

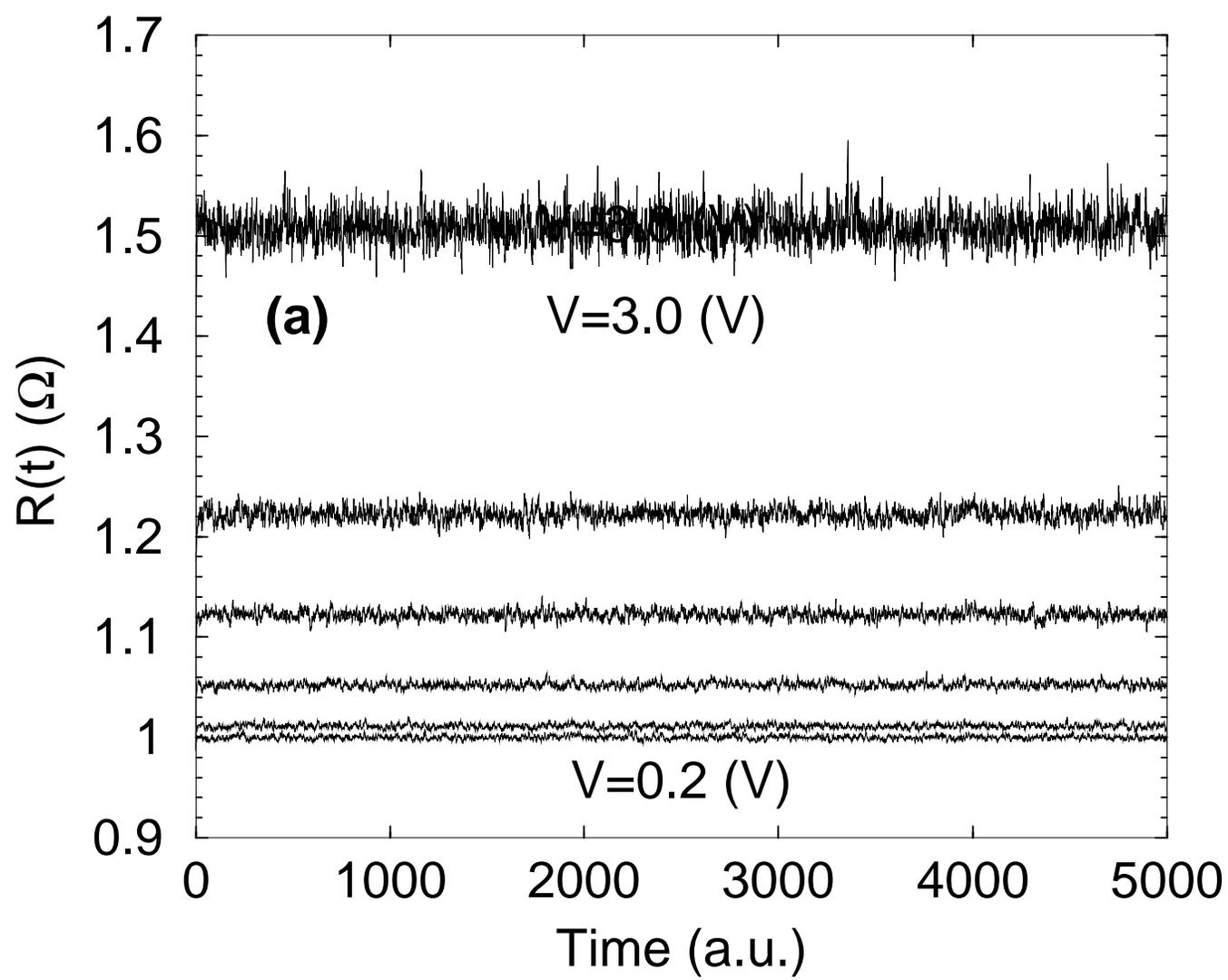
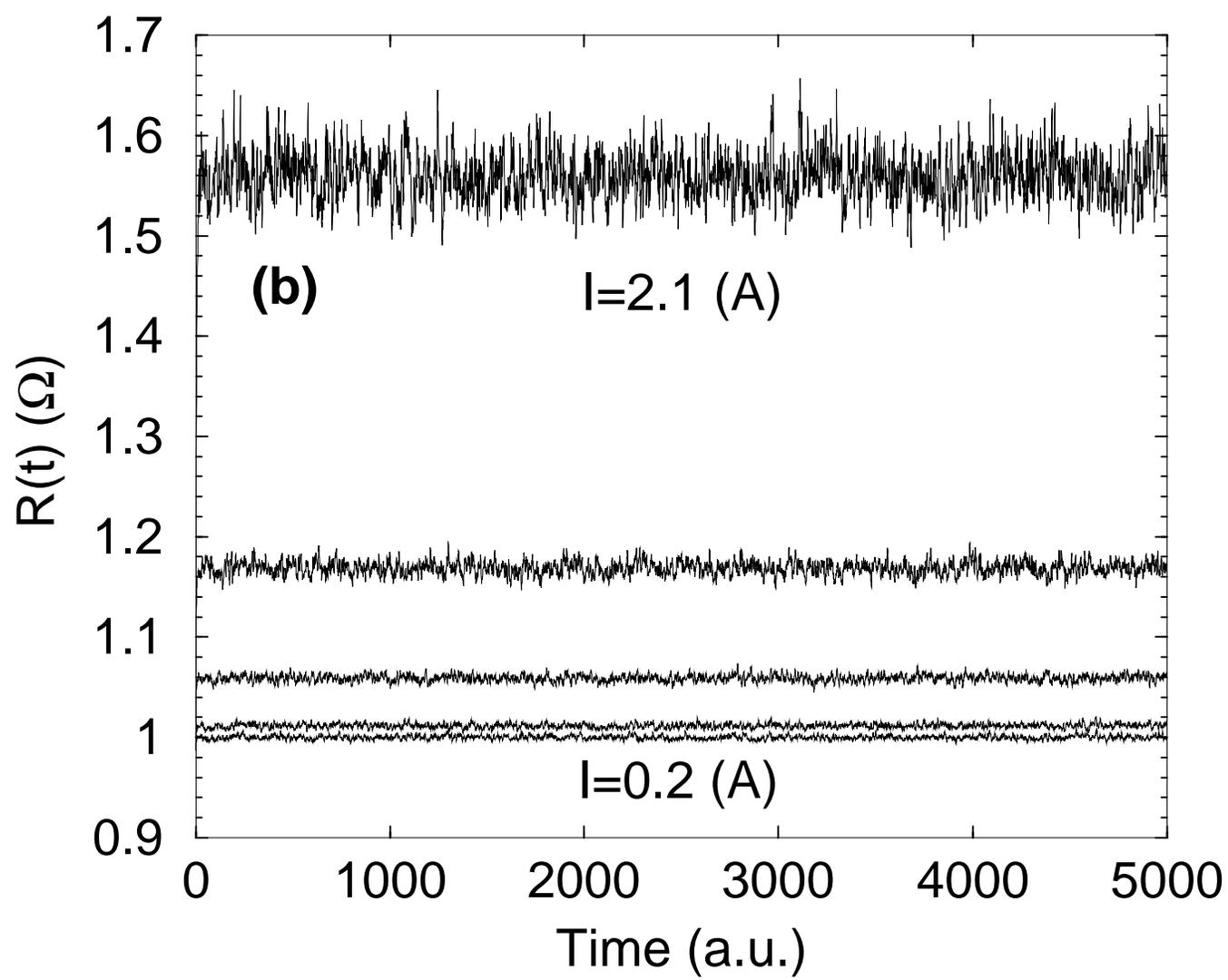
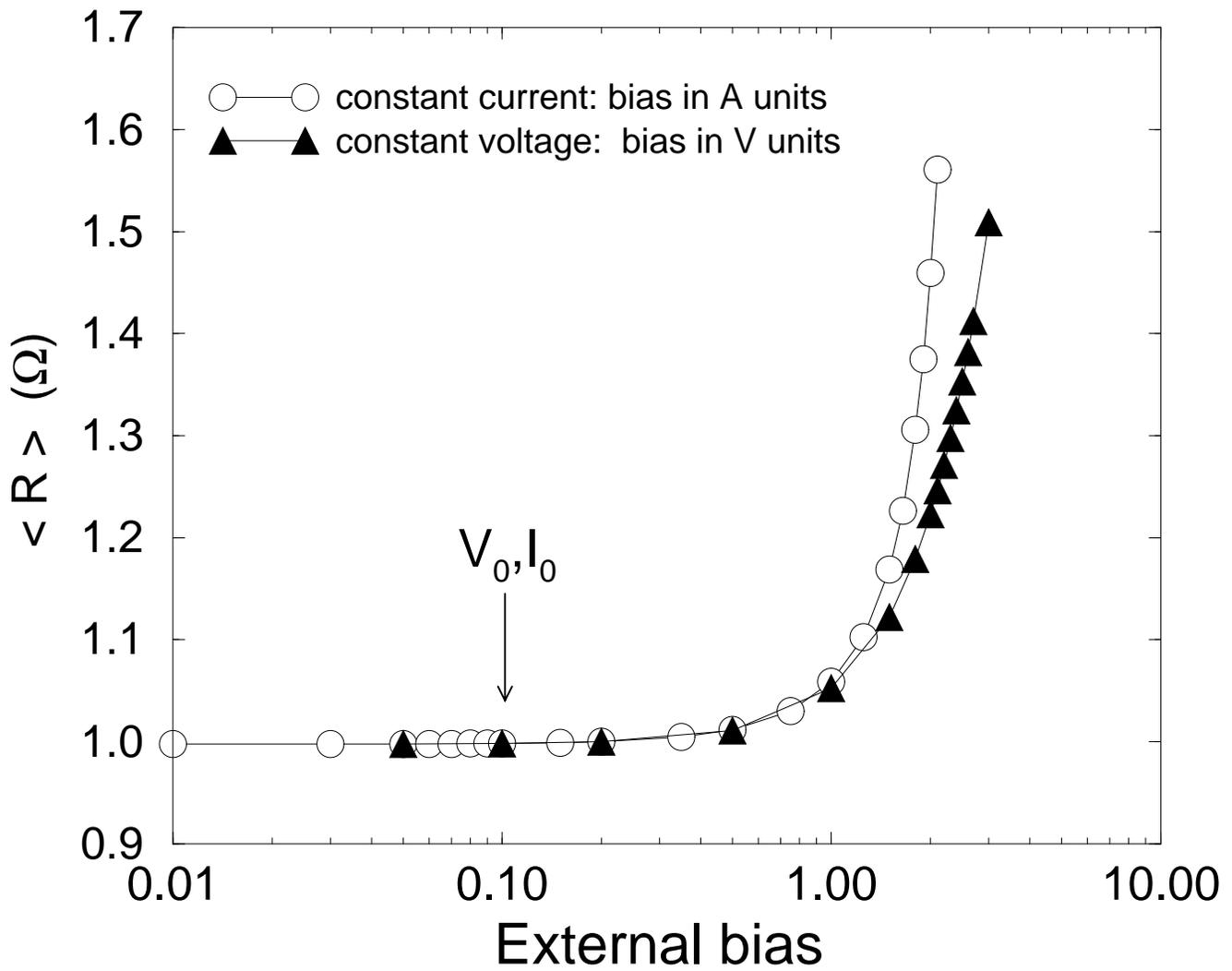
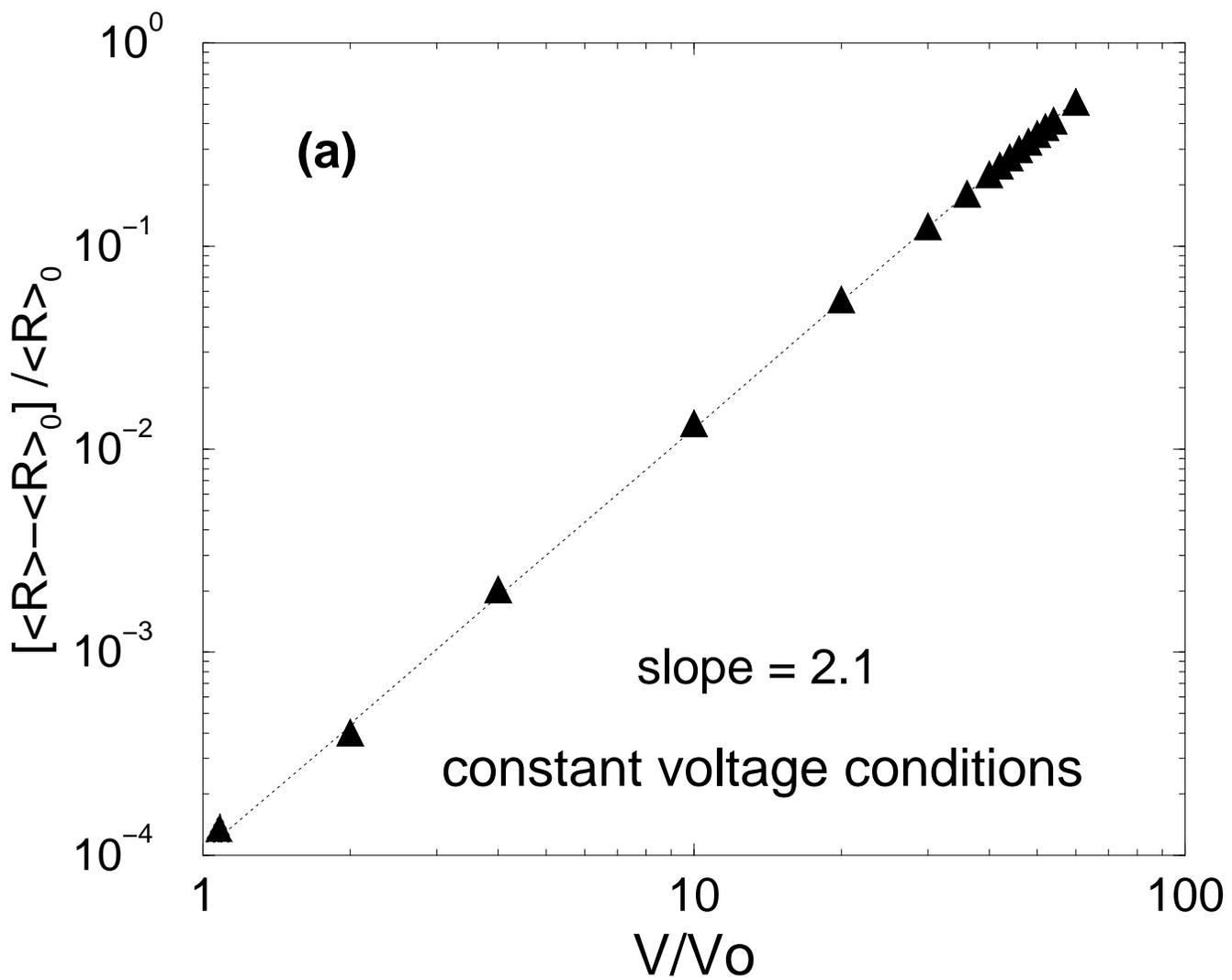
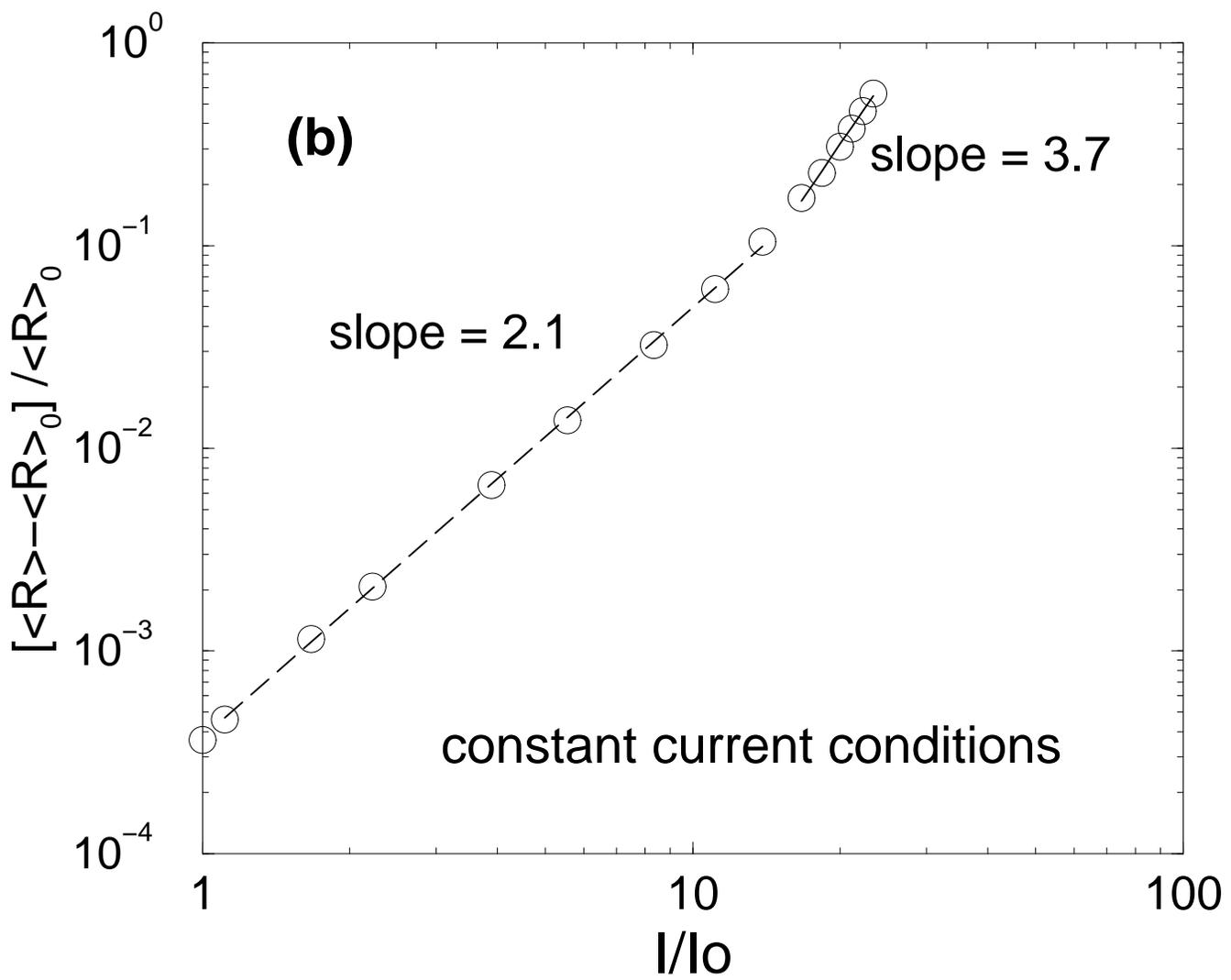
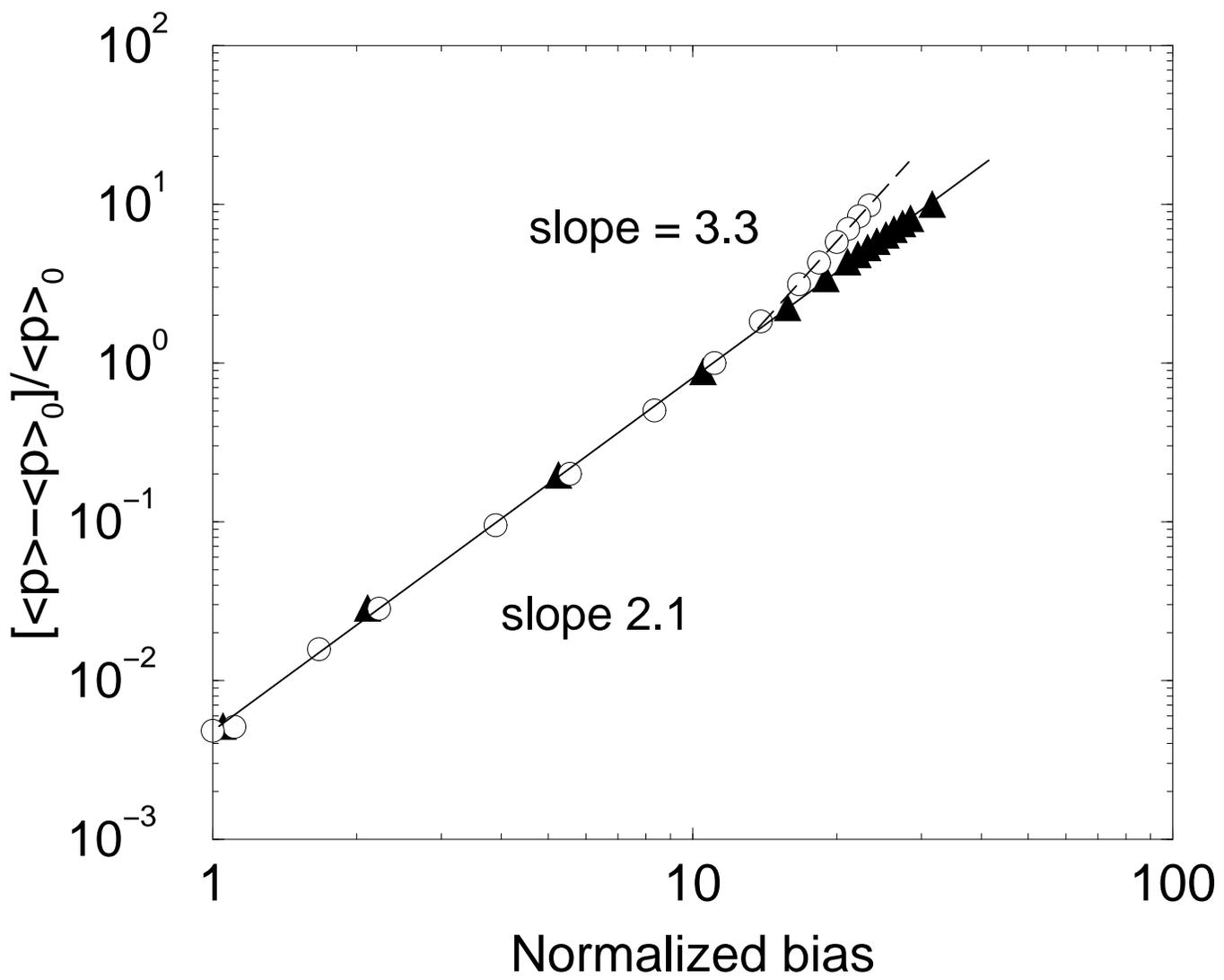
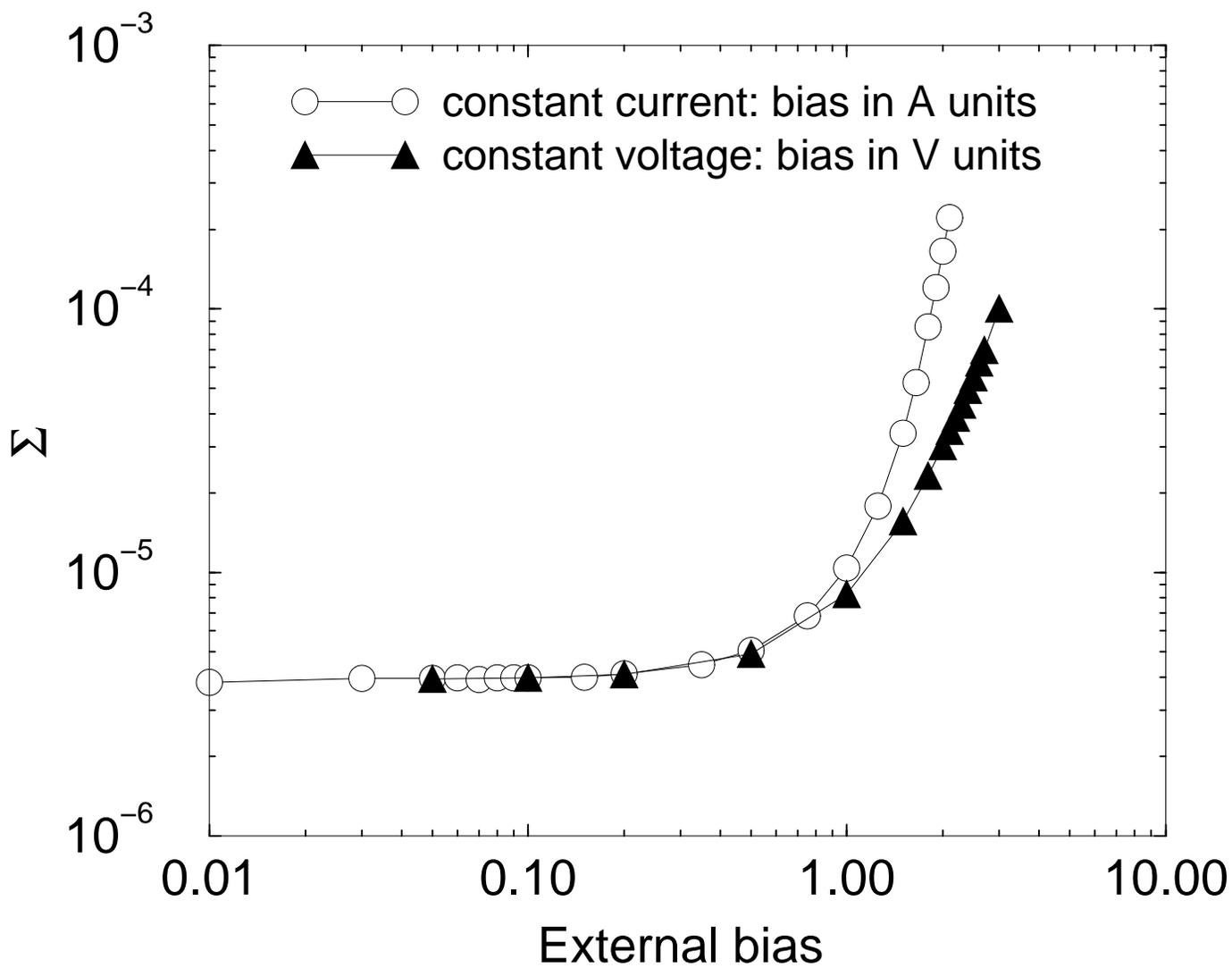
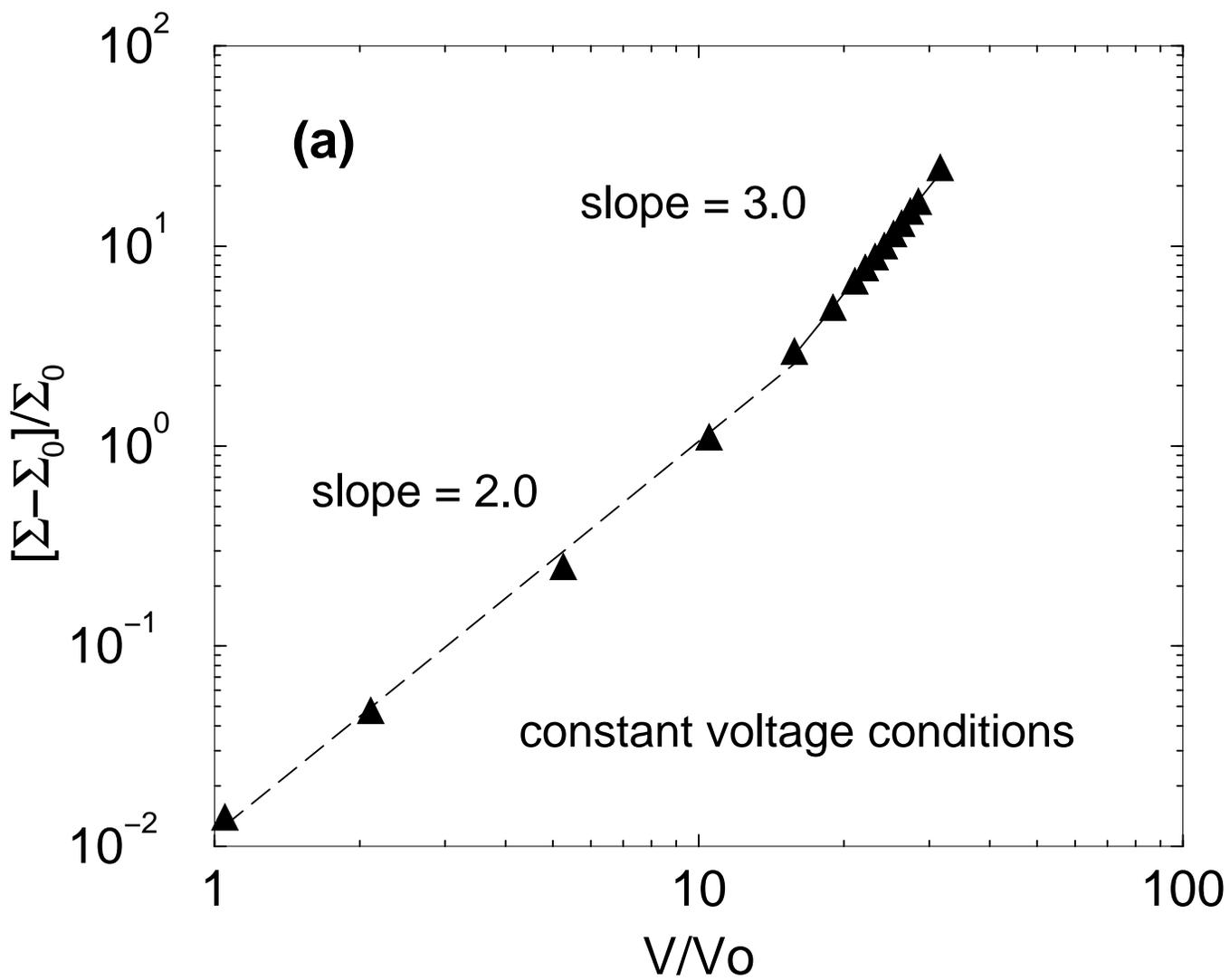
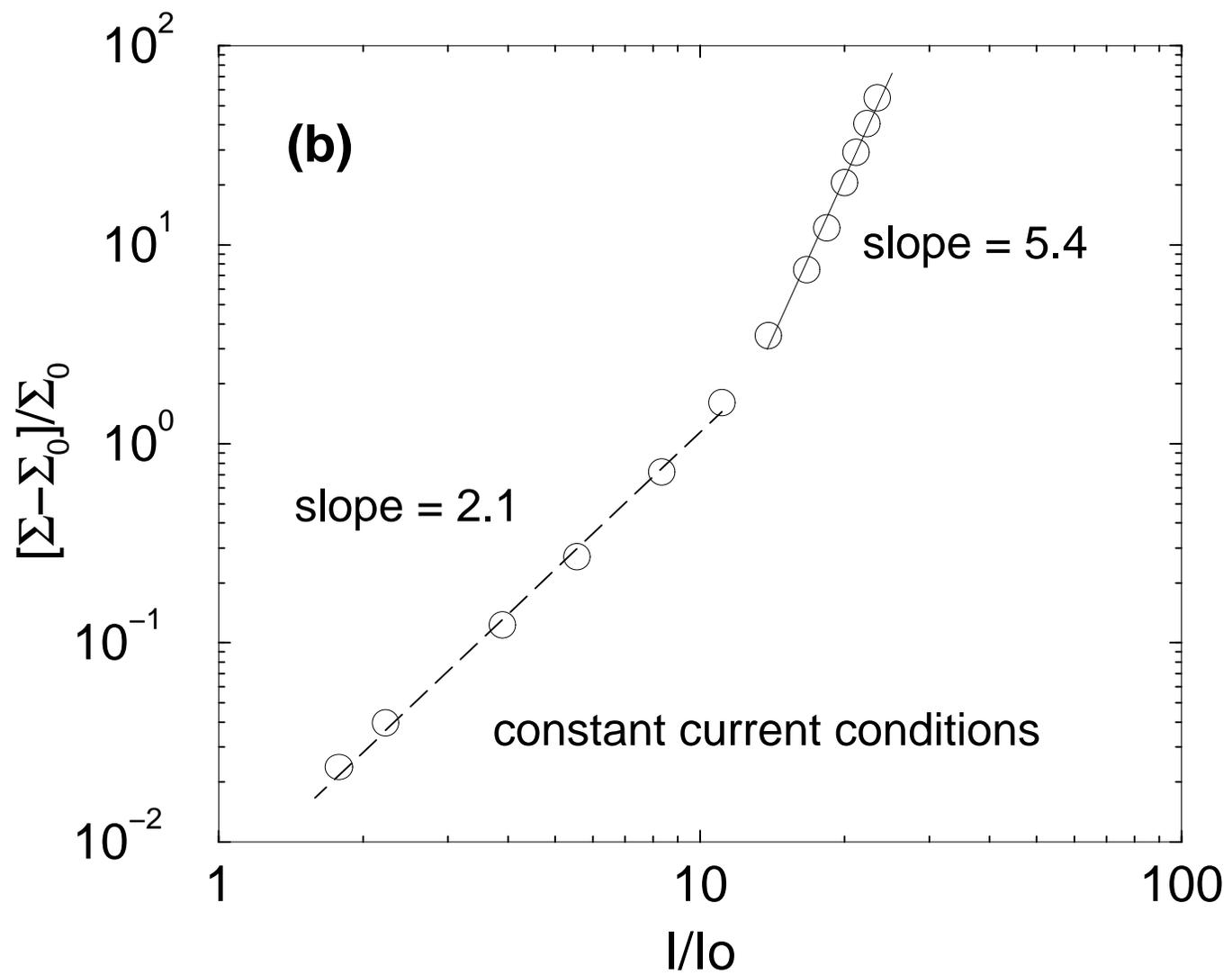
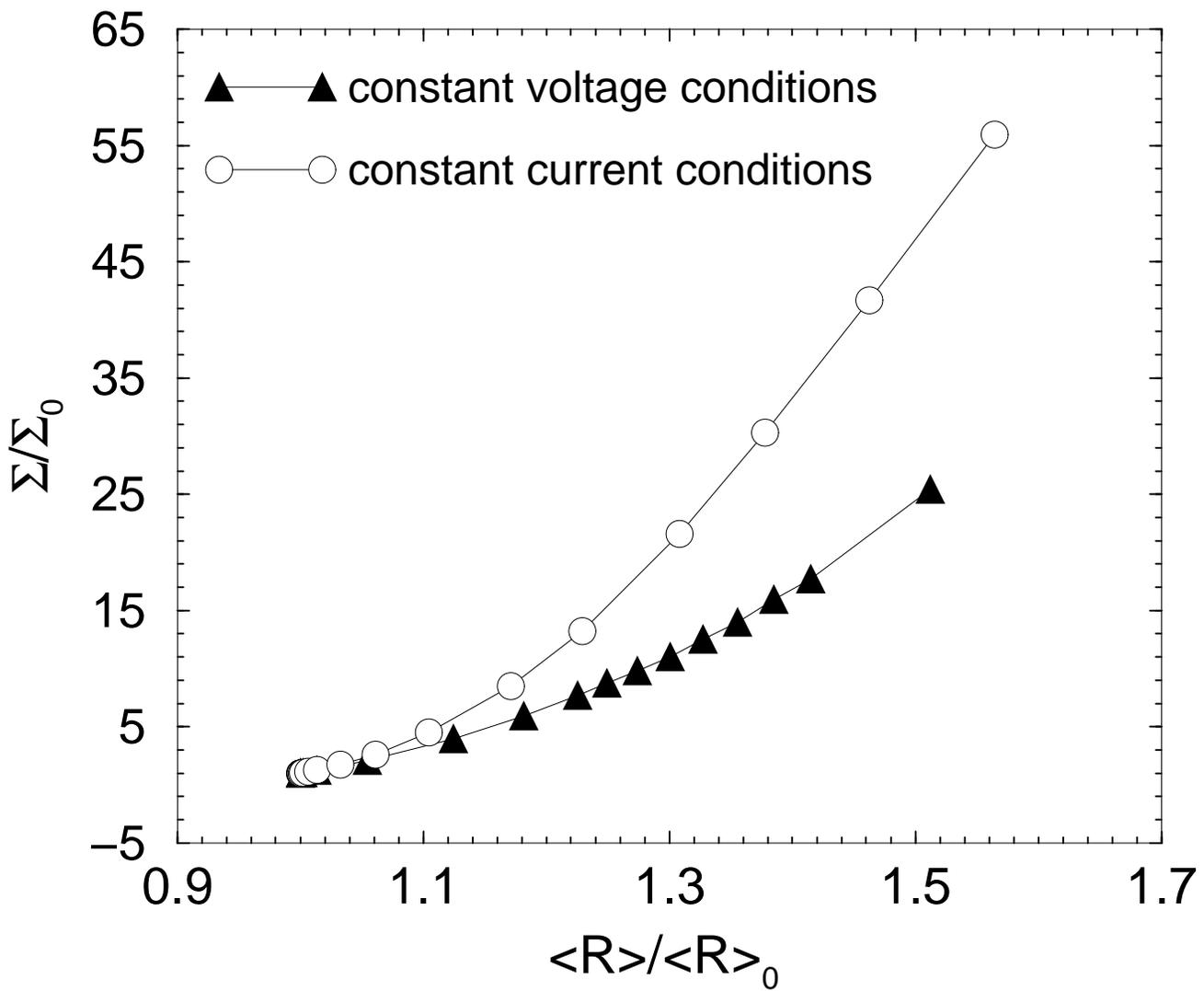
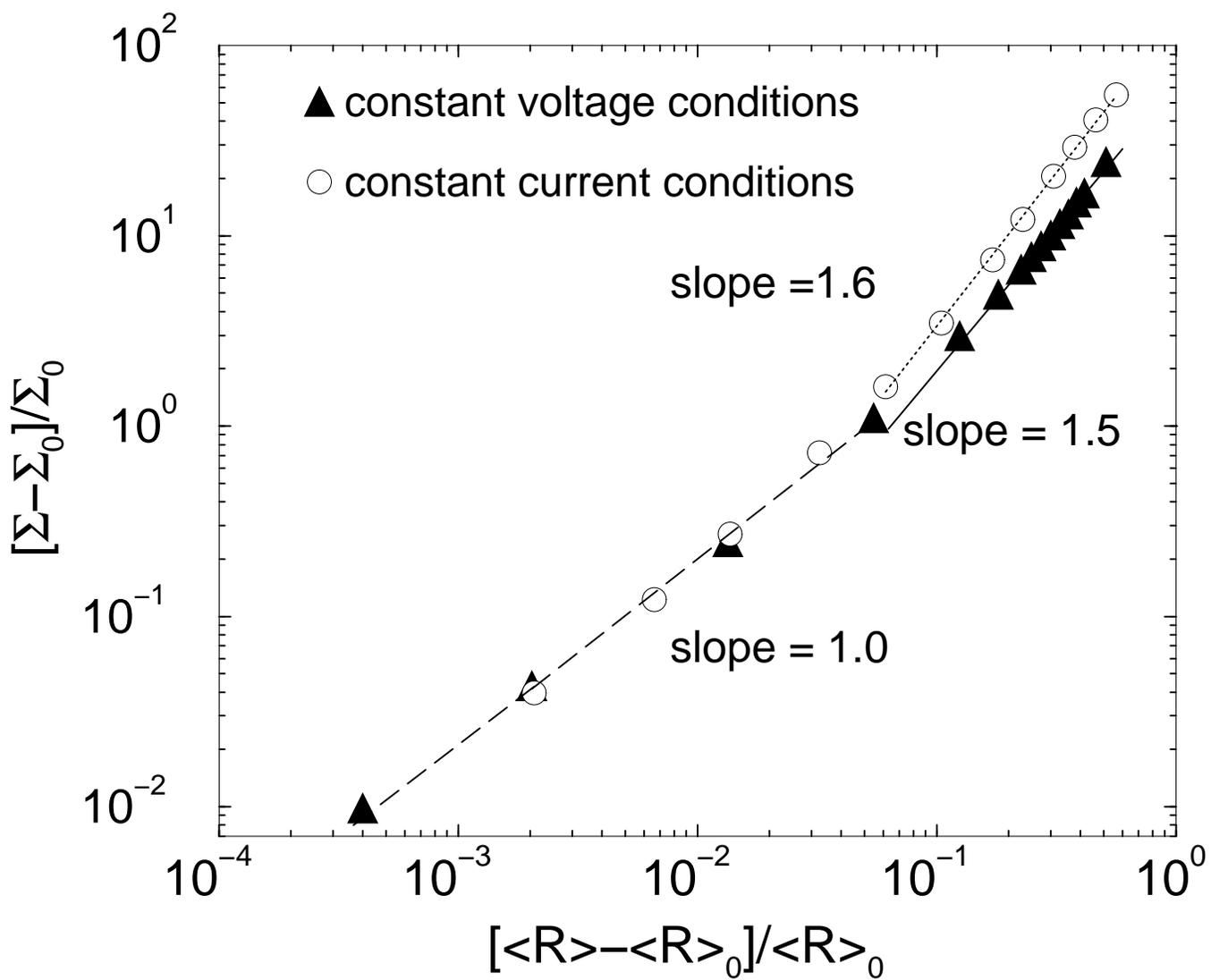
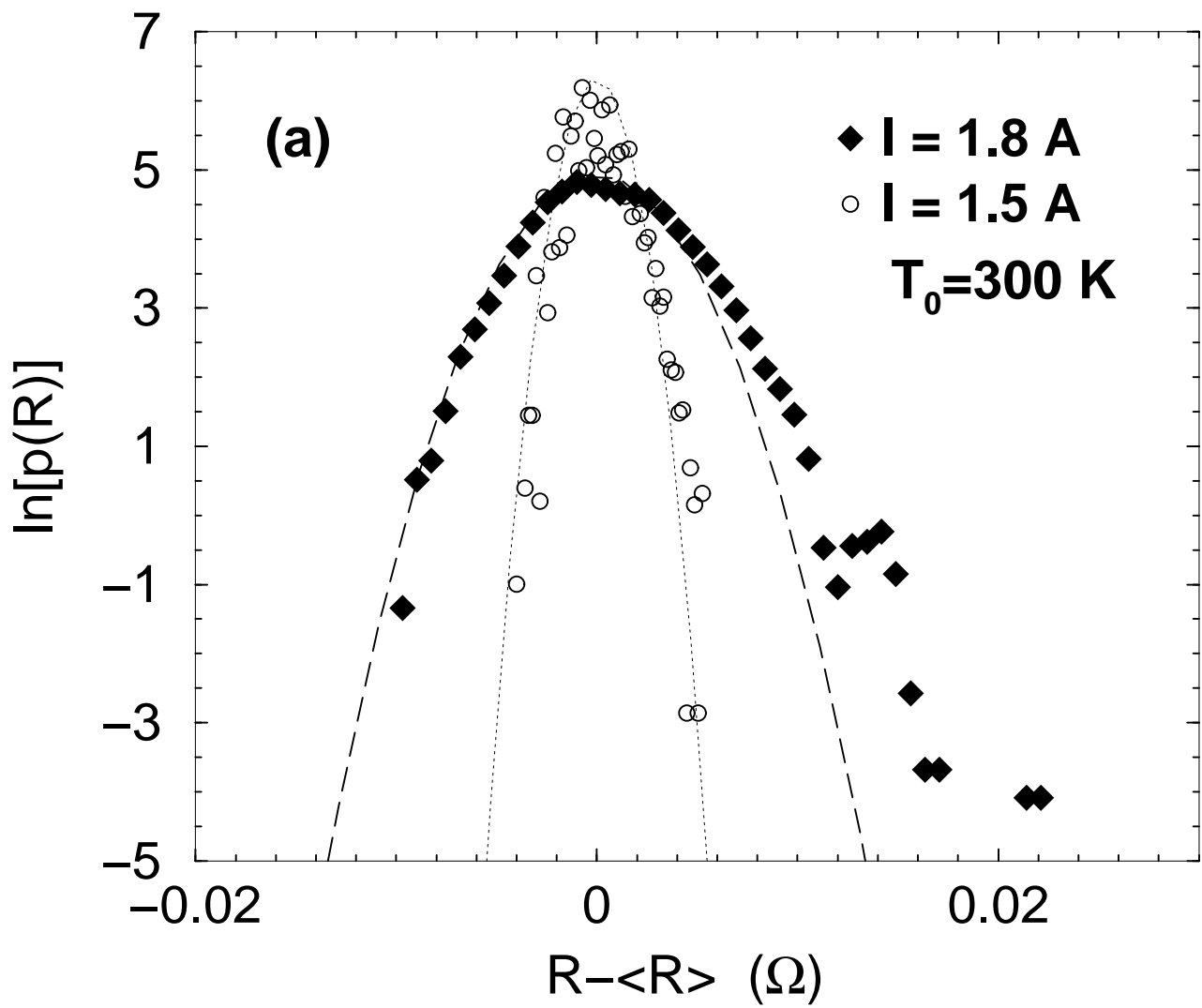
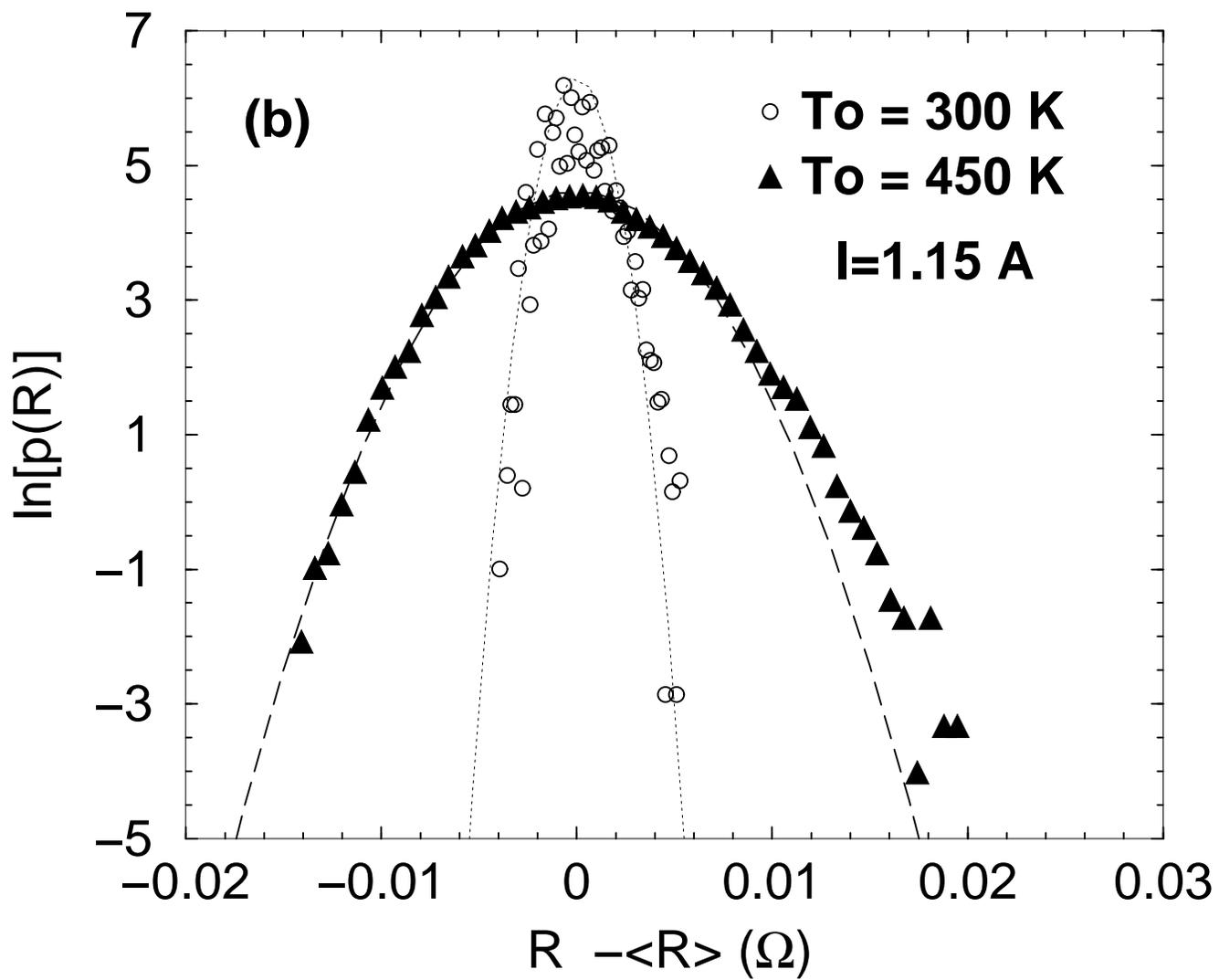
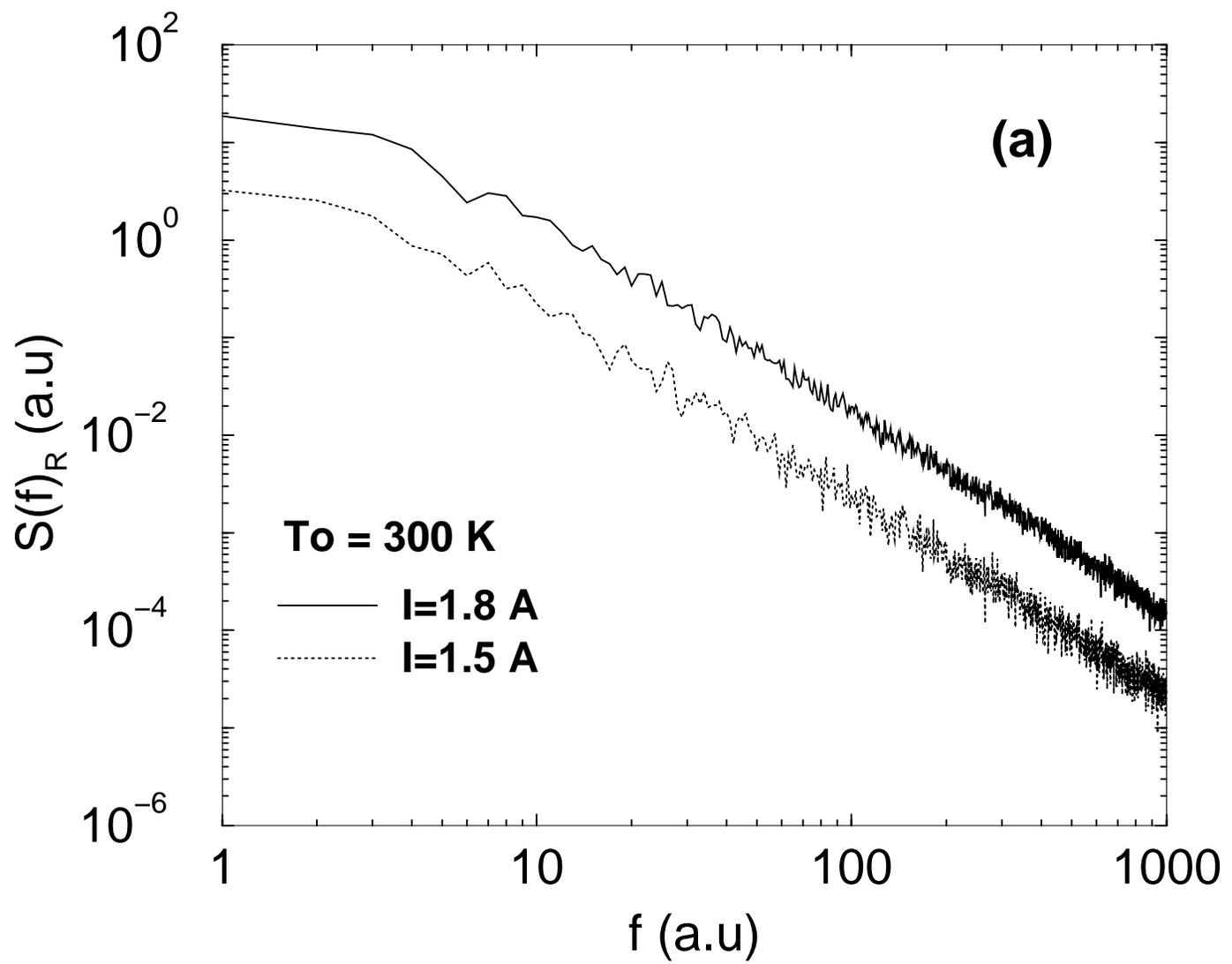
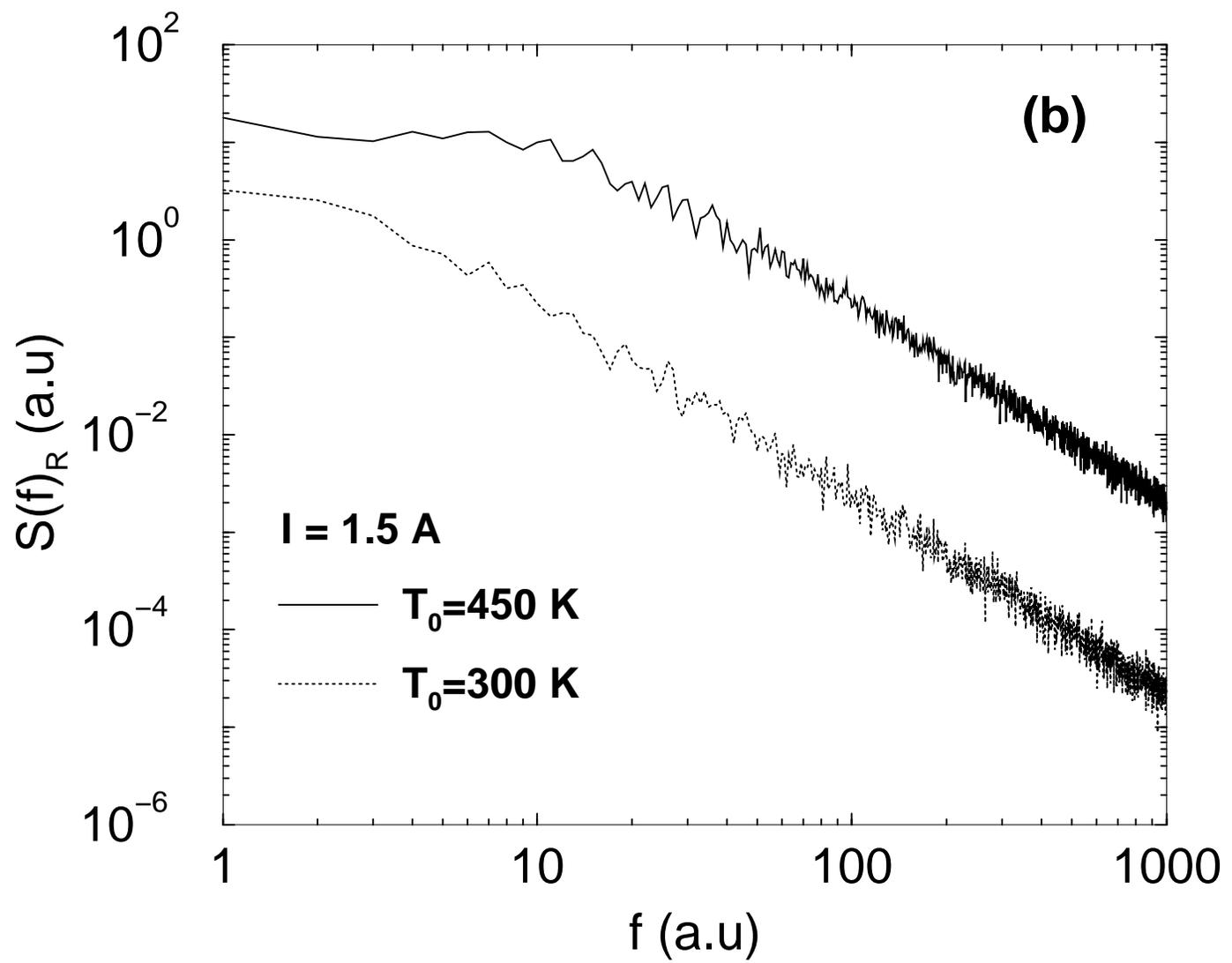
\end{document}